\documentclass[aps,preprint]{revtex4}%
\usepackage{amsfonts}
\usepackage{amsmath}
\usepackage{amssymb}
\usepackage{graphicx}%
\usepackage{pstricks}
\usepackage{caption}
\begin{document}
\preprint{ }
\title{Thermal frequency noise in low oscillation amplitude Dynamic Scanning Force Microscopy}
\author{J. Colchero, M. Cuenca, J.F. Gonz\'{a}lez Mart\'{\i}nez, J. Abad, E.
Palacios-Lid\'{o}n and J. Abell\'{a}n} \affiliation{Facultad de
Qu\'{\i}mica, Departamento de F\'{\i}sica, Universidad de Murcia,
E-30100 Murcia.} 

\begin{abstract}
Thermal fluctuation of the cantilever position sets a fundamental limit for
the precision of any Scanning Force Microscope. In the present work we analyse
how these fluctuations limit the determination of the resonance frequency of
the tip-sample system. The basic principles of frequency detection in Dynamic
Scanning Force Microscopy are revised and the precise response of a typical
frequency detection unit to thermal fluctuation of the cantilever is analysed
in detail. A general relation for thermal frequency noise is found as a
function of measurement bandwidth and cantilever oscillation. For large
oscillation amplitude and low bandwidth, this relation converges to the result
known from the literature, while for low oscillation amplitude and large
bandwidth we find that the thermal frequency noise is equal to the width of
the resonance curve and therefore stays finite, contrary to what is predicted
by the relation known so far. The results presented in this work fundamentally
determine the ultimate limits of Dynamic Scanning Force Microscopy.

\end{abstract}
\maketitle

\section{Introduction}

Since its invention almost 20 years ago, Scanning Force Microscopy
(SFM)\cite{SFM} has become an extremely powerful tool for a huge variety of
nanoscale investigations. With respect to resolution and sensitivity, Dynamic
Scanning Force Microscopy (DSFM)\cite{DSF1} seems to be the most promising
technique. Even though (true) atomic resolution was first achieved with
contact mode in liquid environment\cite{atomicLiquid}, now atomic resolution
studies are generally performed with DSFM working in UHV
environment\cite{atomicUHV,atomicSemicond,AtomsOscar}. Recently even
sub-atomic resolution has been reported using DSFM \cite{subatomicUHV}. In
spite of the impressing advances of SFM and DSFM we believe that the ultimate
limit of these techniques is still an open issue. For most applications, the
temperature at which quantum limits become relevant, $T_{Q}=\hbar\omega
_{0}/k\simeq1\mu K$, is well below typical temperature ranges used in SFM.
Correspondingly, either thermal vibration of the
cantilever\cite{DuerigNoiseSTM,thercal,Tesis,ButtNoise}, or fundamental limits
of the detection technique used\cite{Tesis,BeamDeflection,SPMProceedings} -
related to shot noise of the \textquotedblleft detection
particles\textquotedblright\ - determine the resolution in SFM and DSFM. In
most practical applications, the fundamental limit of SFM and DSFM is set by
thermal noise. Thermal noise in an SFM set up is the consequence of the
equipartition theorem, which relates the mean energy of the cantilever with
the thermal energy $kT$,
\begin{equation}
\frac{1}{2}c~a_{th}^{2}=\frac{1}{2}kT\label{Equipartition}%
\end{equation}
where $a_{th}^{2}$ is the (mean square) displacement of the cantilever induced
by thermal fluctuation, and $c$ is its force constant. For high resolution
distance measurements stiff
cantilevers\cite{GiessiblOptimumParameters,GiessiblEvaluationTF} should be
used ($\delta z=\sqrt{kT/c}$, with $\delta z$ fluctuation of tip-sample
distance), while for high resolution force measurements soft
cantilevers\cite{RugarSoft} are needed ($\delta F_{th}=c\cdot a_{th}%
=\sqrt{c~kT}$). Although the mean displacement and the mean force fluctuation
of the cantilever are important quantities, in many applications they do not
directly determine resolution, either because appropriate filtering
significantly reduces the measured noise, or because the SFM technique used
-as for example DSFM- is not directly limited by the displacement or the force.

In typical DSFM applications the cantilever is excited at or near its natural
frequency and the variation of its resonant properties -oscillation amplitude,
resonance frequency or quality factor - are recorded. For imaging applications
in liquids or air usually the oscillation amplitude is used as control
parameter that defines constant tip-sample distance\ and the frequency shift
is measured as secondary channel. In UHV applications DSFM operation is the
other way, the frequency shift is the control parameter for tip-sample
distance and the oscillation amplitude -dissipation energy- is measured as
secondary channel. At the moment DSFM is the most sensitive SFM technique to
measure tip-sample interaction, which is detected as a shift of the resonance
frequency of the system. Therefore, with regard to the ultimate limits of SFM
a key issue -to be discussed in the present work- is to understand in detail
the effect of thermal fluctuations on DSFM detection schemes and in particular
on frequency detection. At present, the thermal noise density of a frequency
measurement is assumed to be\cite{DSFMAlbrecht,DSFMDuerig,RevGiessibl}%
\begin{equation}
\frac{\Delta\nu_{th}}{\nu_{0}}=\sqrt{\frac{kT~}{~\pi~c~a^{2}\ \nu_{0}\ Q}%
}\sqrt{bw} \label{freqNoiseOld}%
\end{equation}
with $Q$ the quality factor, $\nu_{0}$ the resonance frequency, $a$ the (root
mean square!) oscillation amplitude and $bw$ the bandwidth of the measurement.
Note that this relation diverges for small oscillation amplitude and is
proportional to the square root of the measurement bandwidth.

In the present work we will revise in detail how thermal fluctuation limits
the measurement of frequency shift. We find that neither for large bandwidth
measurements nor for low oscillation amplitude relation \ref{freqNoiseOld} is
correct. We present a general relation valid for all ranges of amplitude and
bandwidth that agrees with relation \ref{freqNoiseOld} in the low bandwidth
and large amplitude range. Finally, we present experimental data that
unambiguously proves the validity of the general relation obtained in this work.

\section{Frequency Detection in Dynamic Scanning Force Microscopy}

A thorough discussion of frequency detection and DSFM is out of the scope of
the present work (see the excellent original
works\cite{DSFMAlbrecht,DSFMDuerig} or recent reviews by Garcia and
Perez\cite{RevRuben} and Giessibl\cite{RevGiessibl}), nevertheless we think it
is important to revise some of its basic principles. The focus of this
revision is rather on a comprehensive physical explanation of the technique
than on a profound analysis of the underlying physiscs and statistical
mechanics (see\cite{DSFMDuerig}) or the electronic details of its
implementation(see \cite{DSFMDuerig,Duerig2ndPLL}).

An externally driven SFM cantilever -usually driven by inertial forces- is a
textbook example of a harmonic oscillator. In many aspects SFM and modern
gravitational wave detectors are governed by the same basic principles (see,
for example, \cite{Smith}). As described in detail in appendix A, the dynamic
properties of such a system are described by the quality factor $Q$, by a
driving amplitude $a_{0}$ and by its natural angular frequency $\omega
_{0}=2\pi\nu_{0}=\sqrt{c/m_{eff}}$, with $\nu_{0}$ natural frequency and
$m_{eff}$ effective mass of the cantilever\cite{CommentFreqAndOmega}. In the
steady state regime, the response of the system to a harmonic excitation
$a_{0}\omega_{0}^{2}\cos(2\pi\nu t)$ can be described by a (complex) amplitude
$A(\nu)$, or by two components $X(\nu)$ and $Y(\nu)$, the in-phase and out of
phase components of the oscillation, corresponding to the real and complex
parts of the complex amplitude $A(\nu)$. The time response of the system is
then $a(t)=X(\nu)\cos(2\pi\nu t)+Y(\nu)\sin(2\pi\nu t)=\left\vert
A(\nu)\right\vert \cos\left(  2\pi\nu t+\varphi(\nu)\right)  $ where
$\varphi(\nu)$ is the phase between the driving force and the response. At its
natural frequency $\nu_{0}$ the phase is $-\pi/2$, the (complex) oscillation
amplitude is $A(\nu_{0})=-iY(\nu_{0})=-ia_{0}Q$ and the in-phase component
$X(\nu_{0})$ vanishes.

Figure 1 shows a schematic set-up of the main components used in a DSFM
detection electronics. The multiplication stages together with the filters
essentially calculate the two components $X(\nu)$ and $Y(\nu)$ -relations
\ref{Xcomp} and \ref{Ycomp} in appendix A- from the measured oscillation of
the cantilever $a(t)$. When enabled, the PI-controller of the DSFM detection
electronics adjusts the driving frequency $\nu_{0}$ of the excitation signal
in order to have $X(\nu)=0$. In this way, the system is locked to the natural
frequency of the cantilever, tracks this frequency if it varies due to
tip-sample interaction and generates an output proportional to the shift of
the resonance frequency. In addition to the main DSFM components figure 1
shows the signals (in the frequency domain) along the different points in the
DSFM detection path. A key component of any DSFM detection unit is the voltage
(VCO) or numerically (NCO) controlled oscillator that generates the excitation
and reference signals for the lock-in detection scheme. In most applications
this oscillator drives the piezo that induces motion of the cantilever. The
corresponding deflection $a(t)$ is analyzed by multiplication with two
reference signals in quadrature.\ We will first assume the most general case
where the frequency $\nu$ of the cantilever motion and that of the reference
oscillator are different. Then, multiplication of the deflection signal
$a(t)=a_{0}\cos(2\pi\nu t)$ with the reference signal $a_{ref}(t)=\cos(2\pi
\nu_{ref}t)$ results in two quadrature signals $x_{q}(t)=a(t)\cos\left(
2\pi\nu_{ref}t\right)  $ and $y_{q}(t)=a(t)\sin\left(  2\pi\nu_{ref}t\right)
$ with frequency components at $\nu_{\Delta}=\nu-\nu_{ref}$, and $\nu_{\Sigma
}=\nu+\nu_{ref}$. As shown in appendix A, $x_{q}(t)$ and $y_{q}(t)$ can be
obtained from $X(\nu)$ and $Y(\nu)$ by multiplication with ($\mathbf{M}%
_{\Delta}(t)+\mathbf{M}_{\Sigma}(t)$), where $\mathbf{M}_{\Delta}(t)$ is a
rotation matrix generating a clockwise rotation with frequency $\nu_{\Delta}$,
and $\mathbf{M}_{\Sigma}(t)$ is a rotation matrix generating a
counter-clockwise rotation with frequency $\nu_{\Sigma}$. In the frequency
domain the signal $a(t)$ is thus splitted and shifted to the sum and to the
difference frequency (see figure 1). After the multiplication stages the two
quadrature signals $x_{q}(t)$ and $y_{q}(t)$ are low-pass filtered over a
timespan proportional to the time constant $\tau$ of the filter. The precise
time domain signals are determined in appendix A. Again, two matrices
$\mathbf{M\tau}_{\Delta}(t)\ $and $\mathbf{M\tau}_{\Sigma}(t)$ can be defined
(relations \ref{MtauDel} and \ref{MTauSum} in appendix A), corresponding to a
clockwise and counter-clockwise rotation with the frequencies $\nu_{\Delta}$
and $\nu_{\Sigma}$. As compared to the first matrices, these ``filtered''
matrices have delay angles $\varphi_{\Delta}\left(  \tau\right)  $ and
$\varphi_{\Sigma}\left(  \tau\right)  $ as well as multiplicative factors
$1/(1+4\pi^{2}\tau^{2}\nu_{\Delta}^{2})$ and $1/(1+4\pi^{2}\tau^{2}\nu
_{\Sigma}^{2})$.

Usually, in DSFM the signal entering the DSFM detection unit is at the natural
frequency $\nu_{0}$, and the reference is at the same frequency $\nu_{ref}%
=\nu_{0}$. Then multiplication of the input signal with the reference signals
will result in spectra around $\nu=0$ and $\nu=2\nu_{0}$. In this case
$\mathbf{M\tau}_{\Delta}(t)=\mathbf{Id}/2$ , with $\mathbf{Id}$ the identity
matrix, and we find, using relations \ref{MtauDel} and \ref{MTauSum}:%

\begin{align}
\left\langle x_{q}(t)\right\rangle _{\tau}  &  =\frac{X(\nu)}{2}+\frac
{X(\nu)(\cos(2\omega_{0}t)-2\omega_{0}\tau\sin(2\omega_{0}t))}{2\left(
1+4\omega_{0}^{2}\tau^{2}\right)  }+\frac{Y(\nu)(\sin(2\omega_{0}%
t)+2\omega_{0}\tau\cos(2\omega_{0}t))}{2\left(  1+4\omega_{0}^{2}\tau
^{2}\right)  }\\
\left\langle y_{q}(t)\right\rangle _{\tau}  &  =\frac{Y(\nu)}{2}+\frac
{X(\nu)(\sin(2\omega_{0}t)+2\omega_{0}\tau\cos(2\omega_{0}t))}{2\left(
1+4\omega_{0}^{2}\tau^{2}\right)  }+\frac{Y(\nu)(-\cos(2\omega_{0}%
t)+2\omega_{0}\tau\cos(2\omega_{0}t))}{2\left(  1+4\omega_{0}^{2}\tau
^{2}\right)  }%
\end{align}
These signals can be conveniently represented in the frequency domain (see
fig.1), where the first peak at $\nu=0$ is one-sided (since no negative
frequency exists) and the second at $\nu=2\nu_{0}$ is two-sided. In typical
DSFM applications the filter is usually set so that $\nu_{0}>>1/\tau$, then it
blocks the $2\nu_{0}$ component and passes the signals from DC to $\nu
_{0}\approx1/\tau$. The value of $\tau$ determines the speed and the
\textquotedblleft cleanness\textquotedblright\ of the signals. Large time
constants (small bandwidth) results in clean but slow response, while,
conversely, small-time constants will result in \textquotedblleft
unclean\textquotedblright\ signals -- in particular with significant $2\nu
_{0}$ component - but with fast response. In our system we have found that
time constants of $3/\nu_{0}$ to $10/\nu_{0}$ give the optimum compromise
between speed and \textquotedblleft cleanness\textquotedblright.

In the frequency domain, the Fourier transforms of the signals $x_{q}(t)$\ and
$y_{q}(t)$ are simply multiplied by the frequency-dependent gain of the
filter. Depending on the time\ constant of the filter, the total amount of
signal may be decreased. The calculation of the total signal $\Delta
u_{bw}\left(  \nu_{c}\right)  $ measured around a frequency $\nu_{c}$ within a
certain bandwidth $bw$ is performed most conveniently in frequency space:%
\begin{equation}
\Delta u_{bw}\left(  \nu_{c}\right)  =\sqrt{%
{\displaystyle\int_{\nu_{c}-bw/2}^{\nu_{c}+bw/2}}
d\nu~\upsilon^{2}(\nu)} \label{signalTotal}%
\end{equation}
where $\upsilon(\nu)$ is the (spectral) signal density (unit: $V/\sqrt{Hz}$)
and $bw=1/\tau$ the effective bandwidth of the filter. The signal densities
corresponding to $\left\langle x(t)\right\rangle _{\tau}$ and $\left\langle
y(t)\right\rangle _{\tau}$ in the frequency domain are $X_{q}(\nu)G_{fil}%
(\nu)$ and $Y_{q}(\nu)G_{fil}(\nu)$, respectively, being $X_{q}(\nu)$ and
$Y_{q}(\nu)$ the Fourier transforms of the quadrature signals $x_{q}(t)$ and
$y_{q}(t)$, and $G_{fil}(\nu)=1/(1+i2\pi\nu\tau)$ the (complex) gain of the
filter (see figure 1).

If the $Q$ factor is low -as in air and in liquids- the signals $\left\langle
x(t)\right\rangle _{\tau}$ and $\left\langle y(t)\right\rangle _{\tau}$ can be
directly used for DSFM. In fact, for low Q factors and low tip-sample
interaction the frequency shift $\Delta\nu_{int}$ induced by tip-sample
interaction is smaller than the width of the resonance ($\Delta\nu_{int}%
<\nu_{0}/Q$). Assuming the validity of the harmonic approximation for the
dynamics of the cantilever, the signal $\left\langle x(t)\right\rangle _{\tau
}$ is then proportional to the frequency shift, and the signal $\left\langle
y(t)\right\rangle _{\tau}$ is proportional to the oscillation amplitude, which
in air and liquids is generally used as control parameter. For high $Q$
factors, however, the width of the resonance is smaller than the frequency
shifts induced by tip-sample interaction. Moreover, high $Q$-factors imply
that the oscillation amplitude needs a long settling time (of the order of
$Q/\nu_{0}$) to reach its steady state value\cite{DSFMAlbrecht}. In this case
it is necessary to track the resonance frequency using Phase Locked Loop
techniques\cite{DSFMAlbrecht,DSFMDuerig,Duerig2ndPLL}. This is implemented
with a PI-controller that essentially adjusts the frequency of the voltage or
numerically controlled oscillator (VCO or NCO, see fig. 1) so that the
$\left\langle x(t)\right\rangle _{\tau}$ component vanishes; the phase of the
oscillation is then kept at $-\pi/2$ and the system is always at resonance.
The output of the PI-controller is then directly proportional to the frequency
shift, and this is the signal used in typical DSFM applications. The
PI-controller represents, from an electronic point of view, a filter with a
well defined bandwidth and gain. The time constant of the filters following
the multiplication stages and the bandwidth and gain of the PI-controller have
to be chosen so that the overall closed loop system is stable\cite{DSFMDuerig}%
. Since the precise set up of the PI-controller does not modify the essential
physics to be discussed here we will assume - to simplify the discussion - an
ideal controller that instantaneously transmits variations of $\left\langle
x(t)\right\rangle _{\tau}$ to the VCO. The time response of the overall system
is then determined by the time constant of the filters after the
multiplication stage, which induce a delay of order $\tau$.

\section{Thermal fluctuation of the cantilever position}

In addition to the \textquotedblleft coherent\textquotedblright\ signal
$a_{ex}(t)$ induced by the external excitation of the cantilever, in the
present context also the \textquotedblleft incoherent\textquotedblright%
\ contribution due to thermal noise, $a_{th}(t)$, is
important\cite{DSFMDuerig}. The total motion of the cantilever is thus
$a(t)=a_{ex}(t)+a_{th}(t)$. When this motion is transduced into an electrical
signal, the position detector will add some instrumental noise $n(t)$, the
total signal entering the DSFM detection unit is then $u_{DSFM}(t)=e~(a_{ex}%
(t)+a_{th}(t))+n(t)$, where $e$ is a constant that describes the sensitivity
(unit:V/nm) of the photodetector electronics. The key question is now: how
does this signal pass the DSFM detection unit and what is the final noise of
the frequency output?

The Equipartition Theorem discussed above (relation \ref{Equipartition})
relates the total thermal displacement to the force constant of the cantilever
and the temperature of the system. However, since the DSFM - detection system
performs non-trivial processing of the input signal, the precise spectrum of
the thermal noise has to be taken into account in order to calculate the noise
of the frequency output and thus the precise limit of DSFM. A detailed
discussion of thermal noise in SFM set-ups is out of the scope of the present
work (see, for example\cite{thercal,ButtNoise,DSFMDuerig,MeyerNoise}). In a
simple picture, the noise density of the cantilever motion can be obtained by
observing that the Equipartition Theorem (equation \ref{Equipartition})
relates the mean energy of the cantilever with the thermal energy $kT$.
Assuming that the effective thermal noise ``force'' driving the cantilever has
a constant spectral noise density $\alpha_{th}$ (unit: $nm/\sqrt{Hz}$), this
coupling strength and the mechanical gain $G\left(  \nu\right)  $ of the
cantilever (relation \ref{gain}) define the spectral response of the thermal
spectrum $a_{th}\left(  \nu\right)  $ of the cantilever: $a_{th}\left(
\nu\right)  =\alpha_{th}$ $G\left(  \nu\right)  $. The Equipartition Theorem
(relation \ref{Equipartition}) then implies:%
\[
\frac{1}{2}kT=%
{\displaystyle\int_{0}^{\infty}}
d\nu~\frac{c}{2}a_{th}^{2}\left(  \nu\right)  =\frac{c}{2}~\alpha_{th}^{2}%
{\displaystyle\int_{0}^{\infty}}
d\nu~\frac{1}{(1-\left(  \nu/\nu_{0}\right)  ^{2})^{2}+((\nu/\nu_{0})/Q)^{2}}%
\]
The second integral gives $\nu_{0}Q~\pi/2$, therefore the thermal noise
density $\alpha_{th}$ is%
\begin{equation}
\alpha_{th}=\sqrt{\frac{2~kT}{\pi~c~\nu_{0}~Q}} \label{densityNoise}%
\end{equation}
The thermal noise density defines the coupling of the thermal bath into the
SFM system. It depends on the quality factor and thus on the dissipation
properties of the system. As the quality factor increases, thermal
fluctuations induce less \textquotedblleft thermal force\textquotedblright\ on
the system, that is, coupling of the thermal bath and the tip-sample system
becomes weaker.

The thermal fluctuation of the cantilever position measured experimentally
with a DSFM detection unit is calculated from the noise density in analogy to
relation \ref{signalTotal}:%
\begin{equation}
\Delta a_{th}\left(  \nu_{c},bw\right)  =\alpha_{th}\sqrt{%
{\displaystyle\int_{\nu_{c}-bw/2}^{\nu_{c}+bw/2}}
d\nu~\left\vert G(\nu)\right\vert ^{2}} \label{thermalNoise}%
\end{equation}
For low frequencies, that is, for frequencies well below the resonance
frequency $G\left(  \nu\right)  =\alpha_{th}$ and the total thermal
fluctuation within a bandwidth $bw$ is $\Delta a_{th}\left(  \nu
_{c},bw\right)  \simeq\alpha_{th}\sqrt{bw}$. At the resonance frequency
$G\left(  \nu\right)  =\alpha_{th}Q$ and the total thermal fluctuation is
\begin{equation}
\Delta a_{th}\left(  \nu_{0},bw\right)  \simeq\alpha_{th}Q\sqrt{bw}%
=\sqrt{\frac{2~kT~Q}{\pi~c~\nu_{0}}}\sqrt{bw} \label{amplitudeNoiseLiterature}%
\end{equation}
as is well known from the literature. While the first relation is usually
correct for static SFM applications, the second requires the condition that
the measurement bandwidth is smaller than the width of the resonance curve:
$bw\simeq1/\tau\ll\nu_{0}/Q$. We note that in general this assumption may not
be correct. In fact, for typical imaging applications (scan speed of 1
second/line, 250 points/line) a minimum bandwidth of 250Hz is required, which
implies $Q\ll4000$ for typical cantilevers ($\nu_{0}\simeq100$ kHz) in order
to fulfill the ``small bandwidth'' approximation. However, in UHV much higher
$Q$ values have been reported\cite{HighQMeyer}. Note that\ the
relation\ \ $\Delta a_{th}\left(  \nu_{0}\right)  =\alpha_{th}Q\sqrt{bw}$
overestimates the thermal noise for large bandwidth, in fact for
$bw>(\pi/2)(\nu_{0}/Q)$ the total thermal noise according to relation
\ref{amplitudeNoiseLiterature} would be larger than $\sqrt{kT/c}$, which is
non-physical since it is in contradiction with the Equipartition Theorem (rel.
\ref{Equipartition}). As discussed below, for a sufficiently large bandwidth
the total thermal noise is $\Delta a_{th}=\sqrt{kT/c}$\ and is independent of
the measuring bandwidth.

A simple illustrative approximation for the spectral noise density of the
tip-sample system is%

\[
\alpha_{appr}(\nu)=\left\{
\begin{array}
[c]{l}%
\sqrt{\frac{Q}{Q+1}}\alpha_{th}~\ \ \text{{\small for} }0\leq\nu<\nu
_{0}\left(  1-\frac{\pi}{4Q}\right)  \ \text{{\small and} }\nu_{0}\left(
1+\frac{\pi}{4Q}\right)  <\nu<\nu_{0}\left(  1+\frac{\pi}{2Q}\right) \\
\sqrt{\frac{Q}{Q+1}}Q~\alpha_{th}~\ \ \text{ {\small for} }\nu_{0}\left(
1-\frac{\pi}{4Q}\right)  <\nu<\nu_{0}\left(  1+\frac{\pi}{4Q}\right) \\
0~\ \ \ \ \ \ \ \ \ \ \ \ \ \text{ {\small for} }\nu\geq\nu_{0}\left(
1+\frac{\pi}{2Q}\right)
\end{array}
\right\}
\]
This approximation shown in figure 2 satisfies as the correct noise density
the relations $\alpha_{appr}(\nu_{0})=Q\;\alpha_{appr}(0)$, width of the peak
of the order $\nu_{0}/Q$, total noise $\Delta a=\left(
{\textstyle\int}
d\nu~\alpha_{appr}^{2}(\nu)\right)  ^{1/2}=\sqrt{kT/c}$ and, for large $Q$,
$\alpha(0)=\alpha_{th}$. The relation of noise in the ``peak'', to that in the
``flat'' part is $\Delta a_{peak}/\Delta a_{flat}=Q$. For high $Q$ factors
most of the noise is in the resonance peak of the curve, and $\Delta
a_{peak}\thickapprox\Delta a_{total}=\sqrt{kT/c}$. Therefore DSFM does not
reduce but rather increase the total thermal noise as compared to static SFM.
However, the signal to noise ratio is not modified: at low frequencies (static
SFM) a driving force $f_{0}$ will induce the motion $a_{0}=f_{0}/c$, and the
signal to noise ratio is $a_{0}/(\alpha_{th}\sqrt{bw})$, while if this force
is applied at resonance it will induce a response $Qa_{0}$ and, for
sufficiently small bandwidth, the thermal noise is $Q\alpha_{th}\sqrt{bw}$.
Noise and signal are therefore amplified equally. Correspondingly, DSFM
increases sensitivity by a factor of $Q$ as compared to static SFM, but the
(theoretical) signal to noise ratio is unchanged. Finally, we note that in
normal applications the thermal fluctuations of the $x$ and $y$ component are
uncorrelated, and that both should have the same amount of fluctuation (see,
however, \cite{RugarNoiseSqueese}). Then, since $\left\langle a_{th}%
^{2}(t)\right\rangle _{\tau}=\left\langle \left\vert x_{th}(t)+i~y_{th}%
(t)\right\vert ^{2}\right\rangle _{\tau}=\left\langle x_{th}^{2}(t)+y_{th}%
^{2}(t)\right\rangle _{\tau}=\left\langle x_{th}^{2}(t)\right\rangle _{\tau
}+\left\langle y_{th}^{2}(t)\right\rangle _{\tau}$, it follows that
$\left\langle x_{th}^{2}(t)\right\rangle _{\tau}=\left\langle y_{th}%
^{2}(t)\right\rangle _{\tau}=\left\langle a_{th}^{2}(t)\right\rangle _{\tau
}/2$, therefore
\begin{equation}
\Delta x_{th}\left(  \nu_{c},bw\right)  =\Delta y_{th}\left(  \nu
_{c},bw\right)  =\Delta a_{th}\left(  \nu_{c},bw\right)  /\sqrt{2}
\label{XYZrelation}%
\end{equation}
and\ for the total thermal fluctuation of the $x$ and $y$ components: $\Delta
x_{th}=\Delta y_{th}=a_{th}/\sqrt{2}=\sqrt{kT/\left(  2c\right)  }$.

\section{Frequency response of DSFM detection schemes to thermal fluctuations}

To calculate the frequency noise, the relation%
\begin{equation}
\Delta\nu_{th}=\left\vert \frac{\partial\nu}{\partial\varphi}\right\vert
\Delta\varphi_{th}=\left\vert \left(  \frac{\partial\varphi}{\partial\nu
}\right)  ^{-1}\right\vert \Delta\varphi_{th} \label{freqNoise}%
\end{equation}
will be used. With $\partial\varphi(\nu_{0})/\partial\nu=-2Q/\nu_{0}$ (see
relation \ref{phase}) the only unknown quantity is the phase noise
$\Delta\varphi_{th}$. The phase, as defined by relation (\ref{phase}) is
$\varphi\left(  t\right)  =-\pi/2+\tan^{-1}(x\left(  t\right)  /y\left(
t\right)  )$. We will assume that DSFM is operated in the Phase Looked Loop
mode and is thus always at resonance, then $\left\langle x(t)\right\rangle
_{\tau}=\left\langle a_{th}\left(  t\right)  /\sqrt{2}\right\rangle _{\tau}=0$
and $\left\langle y(t)\right\rangle _{\tau}=\left\langle Qa_{0}+a_{th}\left(
t\right)  /\sqrt{2}\right\rangle _{\tau}=a_{os}$, with $a_{os}=Qa_{0}$ the
oscillation amplitude at resonance. The correct calculation of the phase noise
is non-trivial, since the phase is a non-linear function of the two variables
$x(t)$ and $y(t)$, which is non-regular at the origin and thus the common
rules for noise/error propagation have to be applied with care. The correct
calculation based on statistical mechanics is presented in appendix B. Here we
will assume that a finite oscillation is applied to the cantilever in order to
prevent the system to be near the origin of the $\left\{  x(t),y(t)\right\}  $
phase space. Then, the mean phase is $\left\langle \varphi(t)\right\rangle
_{\tau}=-\pi/2$\cite{NotePhiIsZero} and the fluctuation of the phase is
$\Delta\varphi_{th}=\sqrt{\left\langle \varphi^{2}(t)\right\rangle _{\tau
}-\left\langle \varphi(t)\right\rangle _{\tau}^{2}}=\sqrt{\left\langle
(\tan^{-1}(x\left(  t\right)  /y\left(  t\right)  ))^{2}\right\rangle _{\tau}%
}$. Using the relation $\left\langle f^{~2}\left(  z_{0}+z\right)
\right\rangle =f^{2}\left(  z_{0}\right)  +\left\vert f^{\prime}\left(
z_{0}\right)  \right\vert ^{2}\left\langle z^{2}\right\rangle $ we have, with
$f\left(  z\right)  =\tan^{-1}\left(  z\right)  $, $z\left(  t\right)
=x\left(  t\right)  /y\left(  t\right)  $ and $z_{0}=\left\langle
x(t)/y(t)\right\rangle =0$,%

\begin{equation}
\Delta\varphi_{th}=\sqrt{\frac{1}{1+\left\langle x(t)/y(t)\right\rangle
_{\tau}^{2}}\left\langle \left(  \frac{x(t)}{y(t)}\right)  ^{2}\right\rangle
_{\tau}}=\sqrt{\frac{\left\langle x^{2}(t)\right\rangle _{\tau}}{\left\langle
y^{2}(t)\right\rangle _{\tau}}} \label{phaseNoise}%
\end{equation}
And finally, with $\left\langle y^{2}(t)\right\rangle _{\tau}=\left(
a_{os}^{2}+a_{th}^{2}/2\right)  $ and relation \ref{XYZrelation} we find for
the phase noise measured around a center frequency $\nu_{c}$ with a bandwidth
$bw=1/\tau$:%

\begin{equation}
\Delta\varphi_{th}=\sqrt{\frac{\Delta a_{th}^{2}\left(  \nu_{c},bw\right)
/2}{a_{os}^{2}+\Delta a_{th}^{2}\left(  \nu_{c},bw\right)  /2}}
\label{phaseNoise2}%
\end{equation}

For large oscillation amplitude, the phase noise is therefore $\Delta
a_{th}/(\sqrt{2}a_{os})$, which can be interpreted as a variation of the phase
due to thermal fluctuation of magnitude $a_{th}/\sqrt{2}$ of the $x$-component
when the $y$-component has an oscillation amplitude $a_{os}$. For very small
oscillation amplitude, relation \ref{phaseNoise2} would give $\Delta
\varphi_{th}=1$. However, as discussed above,\ at the origin of the $\left\{
x(t),y(t)\right\}  $ phase space the phase is mathematically not well defined,
relation (\ref{phaseNoise}) cannot be used, and relation \ref{phaseNoise2} is
not accurate. On physical arguments one would expect a uniform distribution of
phase, that is, a (normalized) probability distribution $p\left(
\varphi\right)  =1/\left(  2\pi\right)  $, which has a mean deviation
$\Delta\varphi=\pi/\sqrt{3}$. As shown in appendix B this is correct in the
limit of vanishing oscillation (see inset of figure 3). The correct relation
for the phase noise has no simple functional relation with the oscillation
amplitude, therefore we propose%
\begin{equation}
\Delta\varphi_{th}=\sqrt{\frac{\Delta a_{th}^{2}\left(  \nu_{c},bw\right)
}{2a_{os}^{2}+3\Delta a_{th}^{2}\left(  \nu_{c},bw\right)  /\pi^{2}}}
\label{phaseNoiseBetter}%
\end{equation}
as approximation for the correct phase fluctuation which has the correct large
and low oscillation behavior. Figure 3 shows the known large oscillation
behavior for phase noise, the correct relation calculated in the appendix as
well as\ the approximations according to relations\ \ref{phaseNoise} and
relation \ref{phaseNoiseBetter} with the correct small and large oscillation limits.

With relation \ref{freqNoise} we finally obtain for the total frequency
noise:
\begin{equation}
\Delta\nu_{th}=\frac{\nu_{0}}{2Q}\sqrt{\frac{\Delta a_{th}^{2}\left(  \nu
_{c},bw\right)  }{2a_{os}^{2}+3\Delta a_{th}^{2}\left(  \nu_{c},bw\right)
/\pi^{2}}} \label{FreqNoiseGeneral}%
\end{equation}
In order to discuss this relation, and to compare with the results known from
the literature, we will consider the different approximations for large and
low oscillation amplitudes, as well as for small and large bandwidth. For
(very) low bandwidth, and large oscillation amplitude we obtain, using
relation \ref{amplitudeNoiseLiterature}%
\[
\Delta\nu_{th}=\frac{\nu_{0}}{2Q}\sqrt{\frac{1~}{\pi}\frac{kT}{ca_{ex}^{2}%
}\frac{Q}{\nu_{0}}}\sqrt{bw}%
\]
which is similar to the result reported in the
literature\cite{DSFMAlbrecht,DSFMDuerig,RevGiessibl} (relation
\ref{freqNoiseOld}). As discussed in the previous section, this (very) ``low
bandwidth approximation'' is usually not valid. On the contrary, we believe
that in most applications the bandwidth is larger than the width of the
resonance curve. Then, as long as instrumental noise is negligible, the
correct approximation would be a ``large bandwidth'' approximation where all
the thermal noise is ``seen'' by the DSFM - detection system. In this case the
corresponding (total) frequency noise is%
\begin{equation}
\Delta\nu_{th}=\frac{\nu_{0}}{2Q}\sqrt{\frac{1}{3/\pi^{2}+2a_{ex}^{2}%
/a_{th}^{2}}} \label{FreqNoiseLargeBw}%
\end{equation}
which is $\Delta\nu_{th}\simeq(\nu_{0}/Q)~a_{th}/(2\sqrt{2}a_{ex})$ for large
amplitude and for $\pi/\left(  2\sqrt{3}\right)  ~\nu_{0}/Q\simeq0.9~\nu
_{0}/Q$ for low amplitude. The characteristic frequency determining the
thermal frequency noise is therefore the width $\nu_{0}/Q$ of the resonance
curve, and for low oscillation amplitudes ($a_{os}\ll a_{th}$) the thermal
frequency noise is essentially given by the width of the resonance curve. In
particular, this implies that, as demonstrated recently for spectroscopy
applications\cite{DSFMThermal}, DSFM is possible without external excitation
of the tip-sample system. Moreover, we believe that a properly designed
DFSM-electronics should be able to lock onto the thermal noise of the cantilever.

\section{Experiments}

In order to confirm the validity of the relations just discussed, noise
measurements have been made as a function of the bandwidth and the oscillation
amplitude. A commercial SFM - system based on optical beam
deflection\cite{nanotec} was used to measure cantilever motion and analysis of
cantilever oscillation was performed either with the DSFM
electronics\cite{nanotec} or with a digital lock-in amplifier\cite{SRI
Instruments}. The set-up of the SFM system and the essential features of the
DSFM electronics are shown in figure 1\cite{CommentAmpContronNo}. A cantilever
with a nominal force constant $c\simeq0.4N/m$\cite{OlympusCantilever} was used
and the tip was kept at a large ($1mm$) distance from the sample. For this
kind of cantilever, relation \ref{Equipartition} gives a total mean (rms)
fluctuation of $100pm$. Figure 4 shows the the spectral noise measurement of
the cantilever movement acquired with the digital locking amplifier\cite{SRI
Instruments}. To characterize the spectral noise density and to discriminate
thermal noise against other (technical) noise sources this data is fitted to
the function
\[
f(\nu)=\frac{\alpha_{th}}{\sqrt{(1-\left(  \nu/\nu_{0}\right)  ^{2})^{2}%
+((\nu/\nu_{0})/Q)^{2}}}+n_{tec}^{0}%
\]
The Lorentzian function is used to describe the thermal noise density of the
cantilever, and the constant $n_{tec}^{0}$ is introduced to describe any
additional (technical) noise (see also \cite{thercal,SPMProceedingsThermal}).
>From the fit to the experimental data, a quality factor $Q=100\pm1$, a
natural frequency $\nu_{0}=79.440\pm0.002kHz$, a thermal noise density
$\alpha_{th}$=26.4$\pm0.2$ fm/$\sqrt{kHz}$ and a constant $n_{tec}^{0}$%
=17$\pm2$ fm/$\sqrt{kHz}$ is found.

The inset of figure 4 shows the total noise as a function of bandwidth, with
the central frequency of the noise measurements at the resonance peak. In this
log-log plot the square root dependence of the total noise on bandwidth for
small bandwidth is clearly recognized from the slope $m=1/2$. For high
bandwidth, the total noise saturates. This saturation occurs for a bandwidth
of the order of the width of the resonance curve $\Delta\nu=\nu_{0}%
/Q\simeq0.8kHz$, in good agreement with the discussion above (relations
\ref{phaseNoise2} and \ref{FreqNoiseLargeBw}) and the data obtained from the
spectral noise density. We note that the saturation of noise is only observed
if other noise sources are negligible, which is clearly the case in our
measurements. Then, as the bandwidth of noise measurement is increased only
the thermal noise in the resonance peak is \textquotedblleft
seen\textquotedblright\ by the DSFM detection unit. If other noise sources are
not negligible, then, as the bandwidth of the DSFM detection unit is increased
($bw>\nu/Q$), the detection unit will ``see'' this additional noise and the
total noise will not saturate. Instead, it will continue to increase with the
square root dependence known from the literature\cite{CommentNoise}. If the
technical noise is appreciable, a total noise well above the theoretical value
$\Delta a_{th}=\sqrt{kT/c}$ for thermal noise can be experimentally observed
since thermal and technical noise is measured.

Finally, figure 5 shows the noise of the frequency as a function of
oscillation amplitude. Two different regimes are recognized: a constant regime
for low oscillation amplitude where the total noise is independent of
oscillation amplitude and a second regime where, as evidenced by the slope
$m=-1$ in the $\log\left(  noise\right)  $ vs. $\log\left(  a_{os}\right)  $
plot, the noise decreases with the inverse of the oscillation amplitude. The
transition range of this graph corresponds to an oscillation amplitude
$\sqrt{kT/c}\simeq100pm$, in good agreement with the value for the thermal
oscillation amplitude obtained from the frequency noise measurement shown in
figure 4.

\section{Summary}

In the present work we have revised DSFM frequency detection and have analyzed
how the thermal fluctuation of the cantilever is processed by a DSFM detection
electronics. We find a general relation for the frequency noise as a function
of a bandwidth and oscillation amplitude. This relation is correct for all
possible values of parameters, while the relation known so far from the
literature is only correct for a particular range. We find that for
sufficiently large bandwidth -that is, small time constants of the DSFM
detection electronics- essentially all the thermal noise of the cantilever is
measured. In this case the width of the resonance peak is the characteristic
noise of any DSFM frequency measurement for small oscillation amplitude, while
for larger oscillation amplitude the noise decreases linearly with the
oscillation amplitude. In the large amplitude and (very) low bandwidth limit
our general relation converges within a constant factor, to the relation known
from the literature.

We are convinced that the results presented in this work are relevant for the
precise determination of the ultimate limits of DSFM. In particular, our
general relation shows that for small oscillation amplitudes the frequency
noise does not diverge, but rather converges towards a finite value.
Therefore, small oscillation DSFM might be much more competitive than
considered up to now. DSFM without external oscillation, that is, driven by
thermal noise, might be possible not only for the measurement of tip-sample
interaction, but also for imaging applications. Since many high precision
measurements SFM measurements\ -in particular in the field of Electrostatic
and Magnetic Force Microscopy- are ultimately based on frequency measurements
we believe that the present work will also improve understanding and
optimization of these related SFM techniques and shed light on the ultimate
limit of Scanning Force Microscopy in different important applications.

\section{Acknowledgments}

The authors acknowledge stimulating discussions with A. Urbina, J. G\'{o}mez,
L. Colchero and A. Gil. The authors also thank Atomic Force F\&E GmbH, and in
particular Mr. Ludger Weisser,\ for supplying the cantilevers\ used. This work
was supported by the Spanish Ministry of Science and Technology through the
projects NAN2004-09183-C10-3 and MAT2006-12970-C02-01.

\section{Appendices}

\subsection{Appendix A}

In the harmonic approximation, the fundamental equation describing the
dynamics of a SFM-system is that of the forced harmonic oscillator,
$m~\ddot{a}(t)+\gamma~\dot{a}(t)~+~c~a(t)~=F(t)$, where $c$ is the force
constant of the system, $m$ its (effective) mass, $\gamma$ the constant
describing the damping in the system and $F(t)$ the external force driving the
oscillator. With the definitions $\omega_{0}=(c/m)^{1/2}$, $Q^{-1}%
=\gamma/(m\omega_{0})=\omega_{0}\gamma/c$, and assuming a harmonic driving
force $F(t)=m~a_{0}\omega_{0}^{2}\cos(\omega t)=m~a_{0}\omega_{0}%
^{2}\operatorname{Re}(e^{i\omega t})$ , where $a_{0}$ is a displacement
determined by the driving force ($a_{0}=F(0)/c$), this equation is transformed
into%
\[
\ddot{a}(t)+(\omega_{0}/Q)~\dot{a}(t)~+~\omega_{0}^{2}a(t)~=a_{0}\omega
_{0}^{2}\cos(\omega t)
\]
Note that, in order to avoid recurrent $2\pi$ factors the angular frequency
$\omega=2\pi\nu$ will be used here instead of the frequency $\nu$ as in the
main text. For the steady state motion this equation can be solved
algebraically with the classical ansatz $a(t)=\operatorname{Re}(A(\omega
)e^{i\omega t})$:%
\[
\operatorname{Re}\left\{  -\omega^{2}A(\omega)~e^{i\omega t}+(\omega
_{0}/Q)~i\omega A(\omega)~e^{i\omega t}+\omega_{0}^{2}~A(\omega)~e^{i\omega
t}\right\}  =\operatorname{Re}(a_{0}\omega_{0}^{2}e^{i\omega t})
\]
from which the complex amplitude $A(\omega)$ is determined as%
\begin{equation}
A(\omega)=\frac{a_{0}}{1-\left(  \omega/\omega_{0}\right)  ^{2}+i(\omega
/\omega_{0})/Q} \label{Amp}%
\end{equation}
A dimensionless (but complex) mechanical gain
\begin{equation}
G(\omega)=\frac{1}{1-\left(  \omega/\omega_{0}\right)  ^{2}+i(\omega
/\omega_{0})/Q} \label{gain}%
\end{equation}
can be defined so that \ $A(\omega)=a_{0}G(\omega)$. For the discussion that
will follow, it is more convenient to describe the complex amplitude
$A(\omega)$ in Cartesian coordinates:
\begin{align}
X(\omega)  &  =a_{0}\frac{1-\left(  \omega/\omega_{0}\right)  ^{2}}{(1-\left(
\omega/\omega_{0}\right)  ^{2})^{2}+((\omega/\omega_{0})/Q)^{2}}%
\label{Xcomp}\\
Y(\omega)  &  =a_{0}\frac{(\omega/\omega_{0})/Q}{(1-\left(  \omega/\omega
_{0}\right)  ^{2})^{2}+((\omega/\omega_{0})/Q)^{2}} \label{Ycomp}%
\end{align}
where $X(\omega)$ and $Y(\omega)$ are the in-phase and out of phase components
of the oscillation. Then the complex amplitude is $A(\omega)=X(\omega
)-iY(\omega)$ and the time response to the driving source $a_{0}\omega_{0}%
^{2}\cos(\omega t)$ is $a(t)=X(\omega)\cos(\omega t)+Y(\omega)\sin(\omega
t)=\left\vert A(\omega)\right\vert \cos\left(  \omega t+\varphi(\omega
)\right)  $ with a phase $\varphi(\omega)$. This phase describes the delay
between the driving source and the response and is
\begin{equation}
\varphi(\omega)=-\tan^{-1}\left(  \frac{Y(\omega)}{X(\omega)}\right)
=-\pi/2+\tan^{-1}\left(  \frac{X(\omega)}{Y(\omega)}\right)  \label{phase}%
\end{equation}
As described in the main text, to experimentally determine state of a harmonic
oscillator, the measured deflection $a(t)$ is multiplied with two reference
signals $\cos\left(  \omega_{r}t\right)  $ and $\sin\left(  \omega
_{r}t\right)  $ to obtain two quadrature signals
\begin{align*}
x_{q}(t)  &  =a(t)\cos\left(  \omega_{r}t\right)  =\frac{X(\omega)}{2}\left(
\cos(\omega_{\Delta}t)+\cos(\omega_{\Sigma}t)\right)  +\frac{Y(\omega}%
{2})\left(  \sin(\omega_{\Sigma}t)+\sin(\omega_{\Delta}t)\right) \\
y_{q}(t)  &  =a(t)\sin\left(  \omega_{r}t\right)  =\frac{X(\omega)}{2}\left(
-\sin(\omega_{\Delta}t)+\sin(\omega_{\Sigma}t)\right)  +\frac{Y(\omega)}%
{2}\left(  \cos(\omega_{\Delta}t)-\cos(\omega_{\Sigma}t)\right)
\end{align*}
with $\omega_{\Delta}=\omega-\omega_{r}$ and $\omega_{\Sigma}=\omega
+\omega_{r}$. With the definitions%
\[
\mathbf{M}_{\Delta}(t)=\frac{1}{2}\left(
\begin{array}
[c]{cc}%
\cos(\omega_{\Delta}t) & \sin(\omega_{\Delta}t)\\
-\sin(\omega_{\Delta}t) & \cos(\omega_{\Delta}t)
\end{array}
\right)  \text{ and }\mathbf{M}_{\Sigma}(t)=\frac{1}{2}\left(
\begin{array}
[c]{cc}%
\cos(\omega_{\Sigma}t) & \sin(\omega_{\Sigma}t)\\
\sin(\omega_{\Sigma}t) & -\cos(\omega_{\Sigma}t)
\end{array}
\right)
\]
the output of the two multiplication stages can be written in matrix notation
as $\left\{  x_{q}(t),y_{q}(t)\right\}  =(\mathbf{M}_{\Delta}(t)+\mathbf{M}%
_{\Sigma}(t))\left\{  X(\omega),Y(\omega)\right\}  $. The corresponding time
evolution can thus be decomposed into one vector rotating clockwise with the
frequency $\omega_{\Delta}$ and another one rotating counter-clockwise with
the frequency $\omega_{\Sigma}$. After the multiplication stage the two
quadrature signals $x_{q}(t)$ and $y_{q}(t)$ are low-pass filtered. For a
simple first order low pass the corresponding time domain signals are%
\begin{align}
\left\langle x_{q}(t)\right\rangle _{\tau}  &  \equiv\frac{1}{\tau}%
{\displaystyle\int_{-\infty}^{0}}
d\xi~x_{q}(t-\xi)~e^{\xi/\tau}\\
&  =\frac{X(\omega)}{2}\frac{\cos(\omega_{\Delta}t)-\omega_{\Delta}~\tau
~\sin(\omega_{\Delta}~t)}{\left(  1+\omega_{\Delta}^{2}\tau^{2}\right)
}+\frac{Y(\omega)}{2}\frac{\sin(\omega_{\Delta}t)+\omega_{\Delta}~\tau
~\cos(\omega_{\Delta}~t)}{\left(  1+\omega_{\Delta}^{2}\tau^{2}\right)
}\nonumber\\
&  +\frac{X(\omega)}{2}\frac{\cos(\omega_{\Sigma}t)-\omega_{\Sigma}~\tau
~\sin(\omega_{\Sigma}~t)}{\left(  1+\omega_{\Sigma}^{2}\tau^{2}\right)
}+\frac{Y(\omega)}{2}\frac{\sin(\omega_{\Sigma}t)+\omega_{\Sigma}~\tau
~\cos(\omega_{\Sigma}~t)}{\left(  1+\omega_{\Sigma}^{2}\tau^{2}\right)
}\nonumber\\
\left\langle y_{q}(t)\right\rangle _{\tau}  &  \equiv\frac{1}{\tau}%
{\displaystyle\int_{-\infty}^{0}}
d\xi~y_{q}(t-\xi)~e^{\xi/\tau}\\
&  =\frac{X(\omega)}{2}\frac{-\sin(\omega_{\Delta}t)-\omega_{\Delta}\tau
\cos(\omega_{\Delta}~t)}{\left(  1+\omega_{\Delta}^{2}\tau^{2}\right)  }%
+\frac{Y(\omega)}{2}\frac{\cos(\omega_{\Delta}t)-\omega_{\Delta}~\tau
~\sin(\omega_{\Delta}~t)}{\left(  1+\omega_{\Delta}^{2}\tau^{2}\right)
}\nonumber\\
&  +\frac{X(\omega)}{2}\frac{\sin(\omega_{\Sigma}t)+\omega_{\Sigma}~\tau
~\cos(\omega_{\Sigma}~t)}{\left(  1+\omega_{\Sigma}^{2}\tau^{2}\right)
}+\frac{Y(\omega)}{2}\frac{-\cos(\omega_{\Sigma}t)+\omega_{\Sigma}~\tau
~\sin(\omega_{\Sigma}~t)}{\left(  1+\omega_{\Sigma}^{2}\tau^{2}\right)  }%
\end{align}
Again, two matrices%
\begin{align}
\mathbf{M\tau}_{\Delta}(t)  &  =\frac{1}{2(1+\tau^{2}\omega_{\Delta}^{2}%
)}\left(
\begin{array}
[c]{cc}%
\cos(\omega_{\Delta}t)-\omega_{\Delta}\tau\sin(\omega_{\Delta}t) & \sin
(\omega_{\Delta}t)+\omega_{\Delta}\tau\cos(\omega_{\Sigma}t)\\
-\sin(\omega_{\Delta}t)-\omega_{\Delta}\tau\cos(\omega_{\Delta}t) &
\cos(\omega_{\Delta}t)-\omega_{\Delta}\tau\sin(\omega_{\Delta}t)
\end{array}
\right)  \text{ }\label{MtauDel}\\
\mathbf{M\tau}_{\Sigma}(t)  &  =\frac{1}{2(1+\tau^{2}\omega_{\Sigma}^{2}%
)}\left(
\begin{array}
[c]{cc}%
\cos(\omega_{\Sigma}t)-\omega_{\Sigma}\tau\sin(\omega_{\Sigma}t) & \sin
(\omega_{\Sigma}t)+\omega_{\Sigma}\tau\cos(\omega_{\Sigma}t)\\
\sin(\omega_{\Sigma}t)+\omega_{\Sigma}\tau\cos(\omega_{\Sigma}t) &
-\cos(\omega_{\Sigma}t)+\omega_{\Sigma}\tau\sin(\omega_{\Sigma}t)
\end{array}
\right)  \label{MTauSum}%
\end{align}
can be defined. The first matrix, $\mathbf{M\tau}_{\Delta}(t)$, corresponds to
a clockwise rotation with the frequency $\omega_{\Delta}$ and a delay angle
$\varphi_{\Delta}=-\tan\left(  \omega_{\Delta}\tau\right)  $\ while the second
matrix, $\mathbf{M\tau}_{\Sigma}(t)$, corresponds to a counter-clockwise
rotation with the frequency $\omega_{\Sigma}$ and a delay angle $\varphi
_{\Sigma}=+\tan\left(  \omega_{\Sigma}\tau\right)  $.

\subsection{Appendix B}

For the calculation of the variation of the phase, we will assume that the
variables $x(t)$ and $y(t)$ are Gaussian variables. At resonance, as discussed
above $\left\langle x(t)\right\rangle _{\tau}=0$ and $\left\langle
y(t)\right\rangle _{\tau}=Qa_{exc}=a_{os}$, therefore their probability
distributions are described by%
\begin{align*}
p_{x}(x)  &  =\sqrt{\frac{1}{\pi a_{th}^{2}}}e^{-(x/a_{th})^{2}}\\
p_{y}(y)  &  =\sqrt{\frac{1}{\pi a_{th}^{2}}}e^{-((y-a_{os})/a_{th})^{2}}%
\end{align*}

These distributions are normalized and have the variance $\Delta x=\Delta
y=a_{th}/\sqrt{2}=\sqrt{kT/(2c)}$. To calculate the distribution of the phase,
first its probability function has to be calculated according to the general
relation (see, for example, \cite{BookStatPhysics}),%
\[
p_{\varphi}(\varphi)=%
{\displaystyle\iint}
dx~dy~p_{x}(x)~p_{y}(y)~\delta(\varphi-\tan^{-1}\left(  x/y\right)  )
\]
which in our case leads to%

\begin{align}
&  =\frac{1}{\pi a_{th}^{2}}%
{\displaystyle\iint}
dx~dy~e^{-(x/a_{th})^{2}}e^{-((y-a_{os})/a_{th})^{2}}~\delta(\varphi-\tan
^{-1}\left(  x/y\right)  )\nonumber\\
&  =\frac{1}{\pi a_{th}^{2}}%
{\displaystyle\iint}
d\vartheta rdr~e^{-(r\sin\left(  \vartheta\right)  /a_{th})^{2}}%
e^{-((r\cos\left(  \vartheta\right)  -a_{os})/a_{th})^{2}}~\delta(\varphi
-\tan^{-1}\left(  r\sin\left(  \vartheta\right)  /r\cos\left(  \vartheta
\right)  \right)  )\nonumber\\
&  =\frac{1}{\pi a_{th}^{2}}%
{\displaystyle\iint}
d\vartheta rdr~e^{-(r^{2}-2r\cos\left(  \vartheta\right)  a_{os}+a_{os}%
^{2})/a_{th}^{2})}~\delta(\varphi-\vartheta)=\frac{1}{\pi a_{th}^{2}%
}e^{-a_{os}^{2}\sin^{2}\left(  \varphi\right)  /a_{th}}\int\limits_{0}%
^{\infty}rdr~e^{-(r-\cos\left(  \varphi\right)  a_{os})^{2}/a_{th}^{2}%
)}\nonumber\\
&  =\frac{1}{\pi a_{th}^{2}}e^{-a_{os}^{2}\sin^{2}\left(  \varphi\right)
/a_{th}^{2}}\left(  \int\limits_{0}^{\infty}dr~(r-\cos\left(  \varphi\right)
a_{os})~e^{-(r-\cos\left(  \varphi\right)  a_{os})^{2}/a_{th}^{2})}%
+\cos\left(  \varphi\right)  a_{os}\int\limits_{0}^{\infty}dr~e^{-(r-\cos
\left(  \varphi\right)  a_{os})^{2}/a_{th}^{2})}\right)  \nonumber\\
&  =\frac{1}{\pi a_{th}^{2}}e^{-a_{os}^{2}\sin^{2}\left(  \varphi\right)
/a_{th}^{2}}\left(  \frac{a_{th}^{2}}{2}~e^{-\cos^{2}\left(  \varphi\right)
a_{os}^{2}/a_{th}^{2})}+\cos\left(  \varphi\right)  a_{os}\frac{\sqrt{\pi}}%
{2}a_{th}\left(  1+\operatorname{Erf}\left[  \frac{a_{os}}{a_{th}}\cos\left(
\varphi\right)  \right]  \right)  \right)  \nonumber\\
&  =\frac{1}{2\pi}\left(  e^{-a_{os}^{2}/a_{th}^{2}}+\sqrt{\pi}\;\frac{a_{os}%
}{a_{th}}\;e^{-a_{os}^{2}\sin^{2}\left(  \varphi\right)  /a_{th}^{2}}%
\cos\left(  \varphi\right)  \left(  1+\operatorname{Erf}\left[  \frac{a_{os}%
}{a_{th}}\cos\left(  \varphi\right)  \right]  \right)  \right)
\label{Phasedistribution}%
\end{align}
\newline where $\operatorname{Erf}[x]=2/\sqrt{\pi}%
{\textstyle\int}
dx~e^{-x^{2}}$ is the normalized Error Function ($\operatorname{Erf}%
[\infty]=1$). This probability distribution is plotted in the inset of figure
3 for the range of oscillation amplitudes $a_{os}/a_{th}=0-2$. For large
oscillation amplitude $a_{os}>>a_{th}$ the first term can be neglected, in
addition we can assume $\operatorname{Erf}[...]\simeq1$, and only very small
angles contribute to the probability amplitude ($\sin\left(  \varphi\right)
\simeq\varphi$; $\cos\left(  \varphi\right)  \simeq1$) then, with
$\varphi_{th}\equiv a_{th}/a_{os}$, we find%

\[
p_{\varphi}(\varphi)\simeq\frac{1}{\sqrt{\pi}\varphi_{th}}e^{-\varphi
^{2}/\varphi_{th}^{2}}%
\]
which is a normalized Gaussian probability distribution of the angle $\varphi$
with variance $\Delta\varphi=\varphi_{th}/\sqrt{2}=a_{th}/(\sqrt{2}a_{os})$,
in agreement with the high excitation limit of eq. \ref{phaseNoise2}. We note
that the ratio $\varphi_{th}\equiv a_{th}/a_{os}$ can be interpreted as the
fluctuation of the phase due to thermal variation $a_{th}/\sqrt{2}$ of the
x-component when the y-component of the oscillation is fixed at $a_{os}$ (for
$a_{os}>>a_{th}$ thermal fluctuation of the y-component essentially gives no
contribution to phase noise). For small oscillation amplitudes the second term
in \ref{Phasedistribution} is small and we find, to first order in
$1/\varphi_{th}$,%

\[
p_{\varphi}(\varphi)\simeq\frac{1}{2\pi}\left( 1+\frac{2\sqrt{\pi}}%
{\varphi_{th}}\cos\left(  \varphi\right) \right)
\]
The angle probability distribution therefore becomes non-Gaussian and
ultimately uniform (see inset of figure 3), as is expected for vanishing
oscillation amplitude. The mean value of this (normalised) angle distribution
is $\overline{\varphi}=0$, and its square deviation $\left(  \Delta
\varphi\right)  ^{2}=\pi^{2}/3-2\sqrt{\pi}\varphi_{th}$. The correct phase
error $\Delta\varphi\left(  a_{os}/a_{th}\right)  $ calculated from the
probability distribution \ref{Phasedistribution} is plotted in figure 3,
together with the different approximations discussed in this work.

\clearpage
\newpage

\section{Figures}

\begin{figure}[Hrb]
\centering
\includegraphics[height=16cm,angle=-90]%
{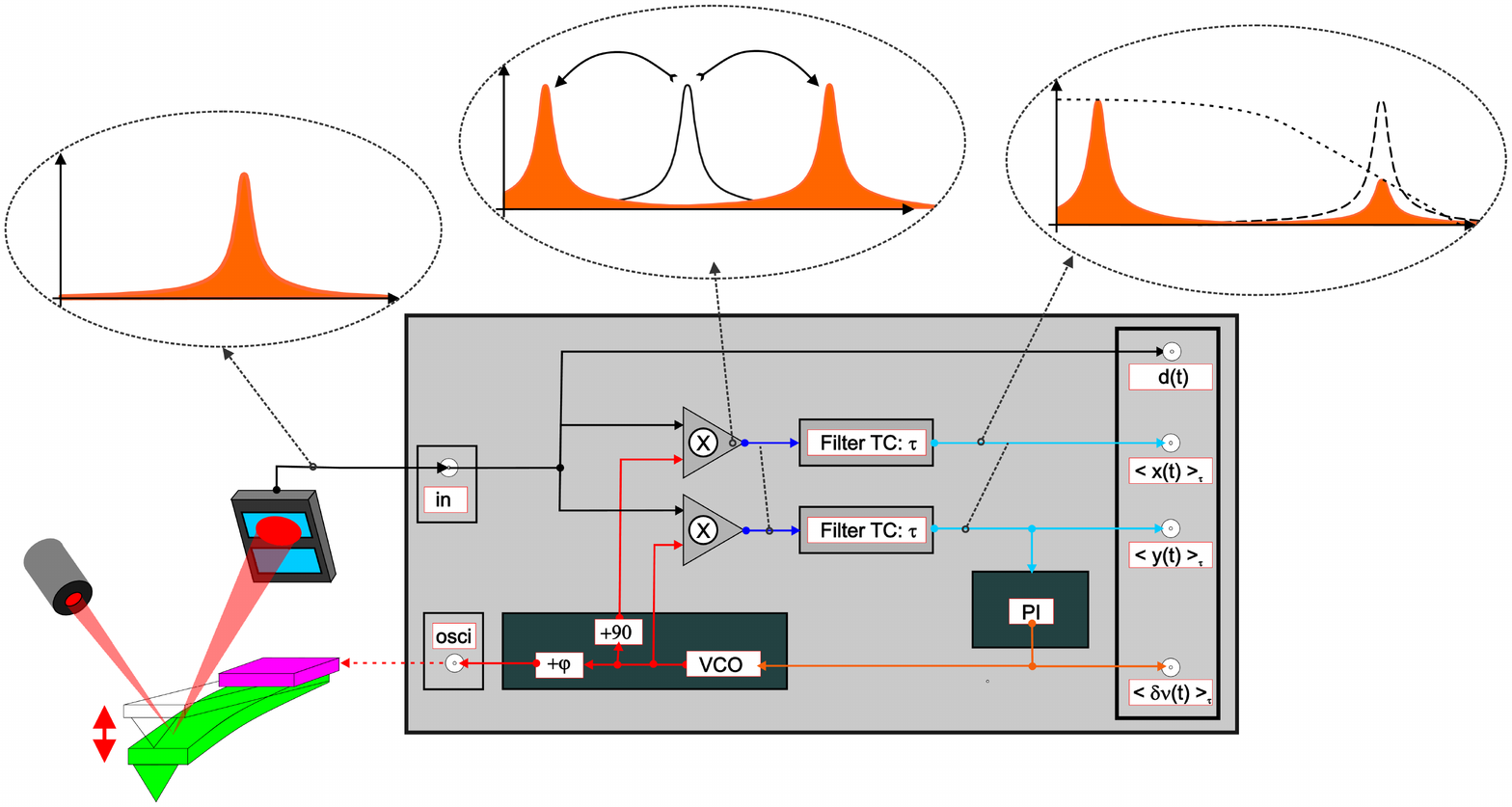}%
\caption{Schematic description of a typical lock-in type DSFM
detection unit. The signal to be analyzed by the DSFM detection is
assumed to be centered around some frequency $\nu_{0}$. It enters
the detection unit at the input ``\textit{in}'', is amplified and
usually high-pass filtered (for simplicity the corresponding
components are not shown) before being multiplied with two reference
signals in quadrature at a frequency $\nu_{ref}$, shifting the
signal to the frequencies $\nu_{0}-\nu_{ref}$ and
$\nu_{0}+\nu_{ref}$. The resulting signals are then low-pass
filtered to remove the higher frequency component
($\nu_{0}+\nu_{ref}$), resulting in two averaged signals
$\left\langle x(t)\right\rangle _{\tau}$ and $\left\langle
y(t)\right\rangle _{\tau}$. For sufficiently small interaction
$\left\langle x(t)\right\rangle _{\tau}$ is proportional to the
frequency shift and can be used to re-adjust the driving frequency
of the VCO (or NCO) by means of an appropriate feedback loop
(PI-controller). The output of the PI-controller used to adjust the
excitation frequency is then proportional to the frequency shift
$\delta \nu(t)$.}
\end{figure}

\begin{figure}[H]
\centering
\includegraphics[height=16cm,angle=-90]%
{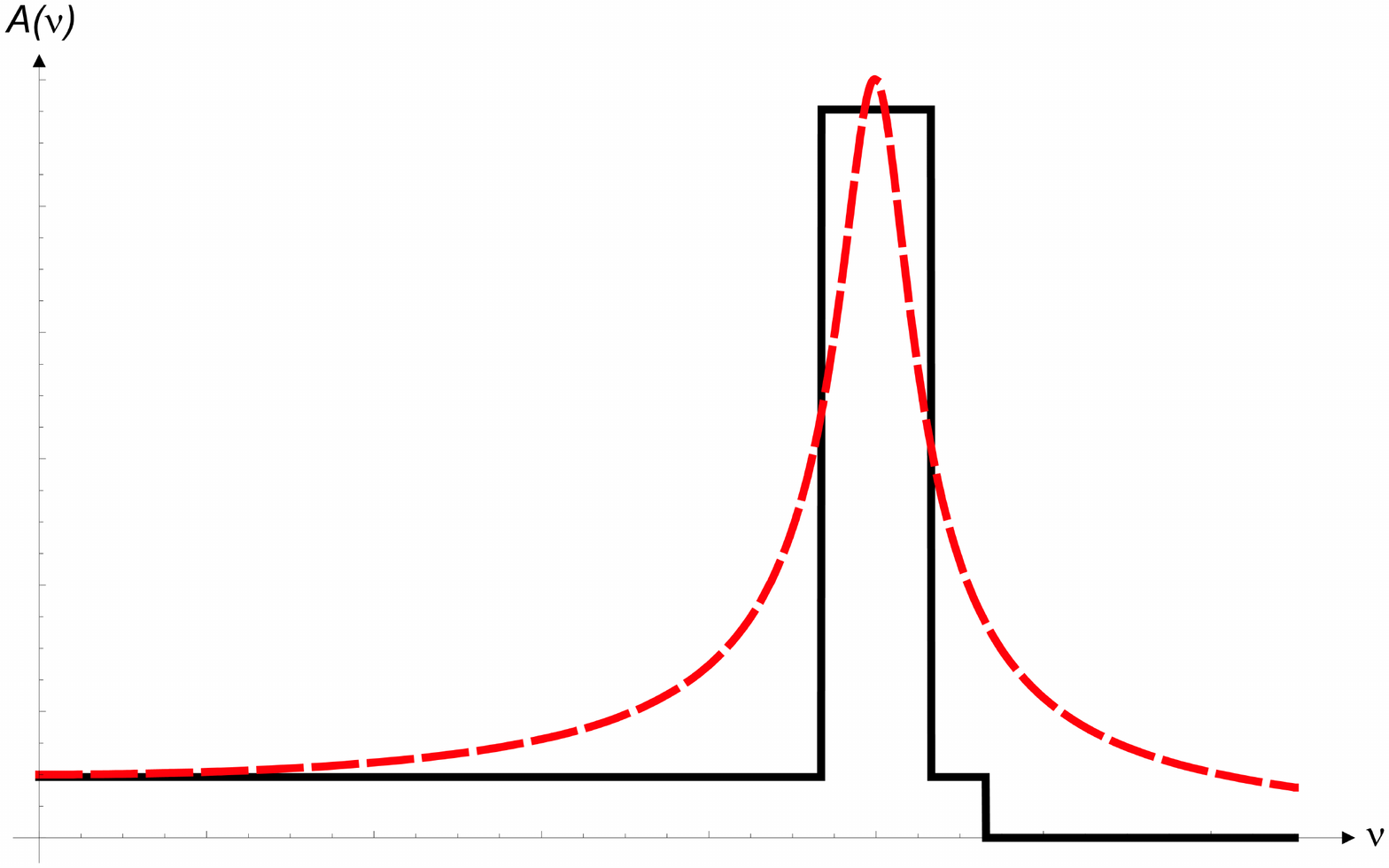}%
\caption{Simple approximation of the noise density (black) for a
thermally excited cantilever as discussed in the main text together
with the correct noise density (red). For large quality factors,
most of the noise is within the main peak at the resonance
frequency.}
\end{figure}

\begin{figure}[H]
\centering
\includegraphics[height=16cm,angle=-90]%
{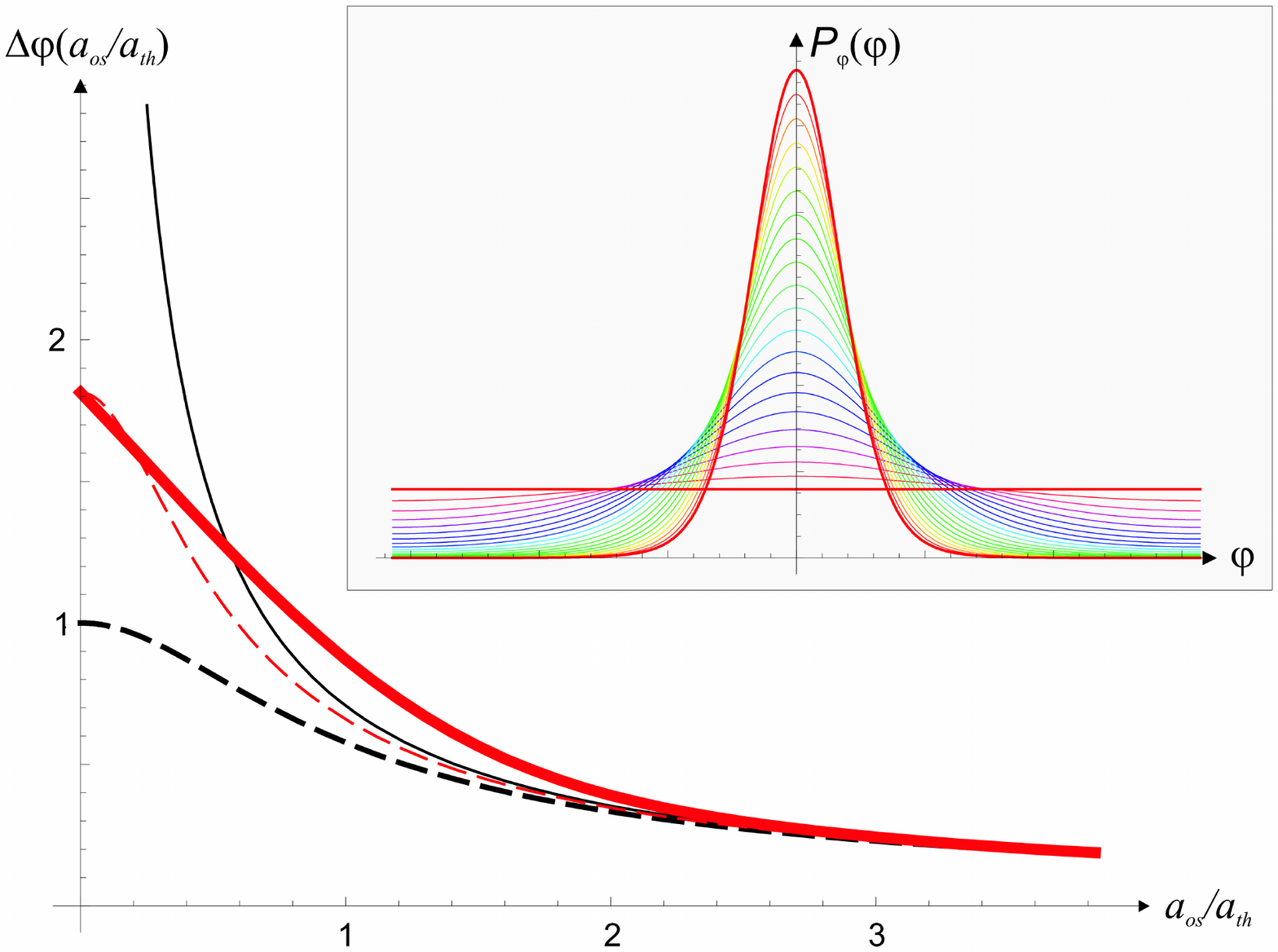}%
\caption{Main graph: Thermal noise error of the phase as a function
of the (relative) oscillation amplitude $a_{os}/a_{th}$. The black,
solid, thin line corresponds to the relation obtained from the
relation known in the literature, which diverges for small
oscillation amplitude. The black, dotted line corresponds
to the relation $\Delta\varphi_{th}=\sqrt{\left\langle (\tan^{-1}%
(x/y))^{2}\right\rangle }$ (\ref{phaseNoise}), which is not correct
at the singular point $\{x,y\}=\{0,0\}$.\ The red, thick, solid line
shows the correct relation calculated from the probability
distribution \ref{Phasedistribution} discussed in the appendix B.
Finally, the red, thin, dotted line corresponds to the approximation
$\Delta\varphi_{th}=\sqrt
{a_{th}^{2}/(2a_{os}^{2}+3a_{th}^{2}/\pi^{2})}$ (relation
\ref{phaseNoiseBetter}), which has the correct low and large
amplitude limits.\newline Inset: Probability distributions
$p_{\varphi}(\varphi)$ for different (relative) oscillation
amplitude $a_{os}/a_{th}$. The probability distributions have been
calculated for the range of oscillation amplitudes
$a_{os}/a_{th}=0-2$. The probability distribution
$p_{\varphi}(\varphi)$ for $a_{os}/a_{th}=0$ is flat while that for
$a_{os}/a_{th}=2$\ is essentially Gaussian and has the highest peak
at $\varphi=0$.}
\end{figure}

\begin{figure}[H]
\centering
\includegraphics[height=16cm,angle=-90]%
{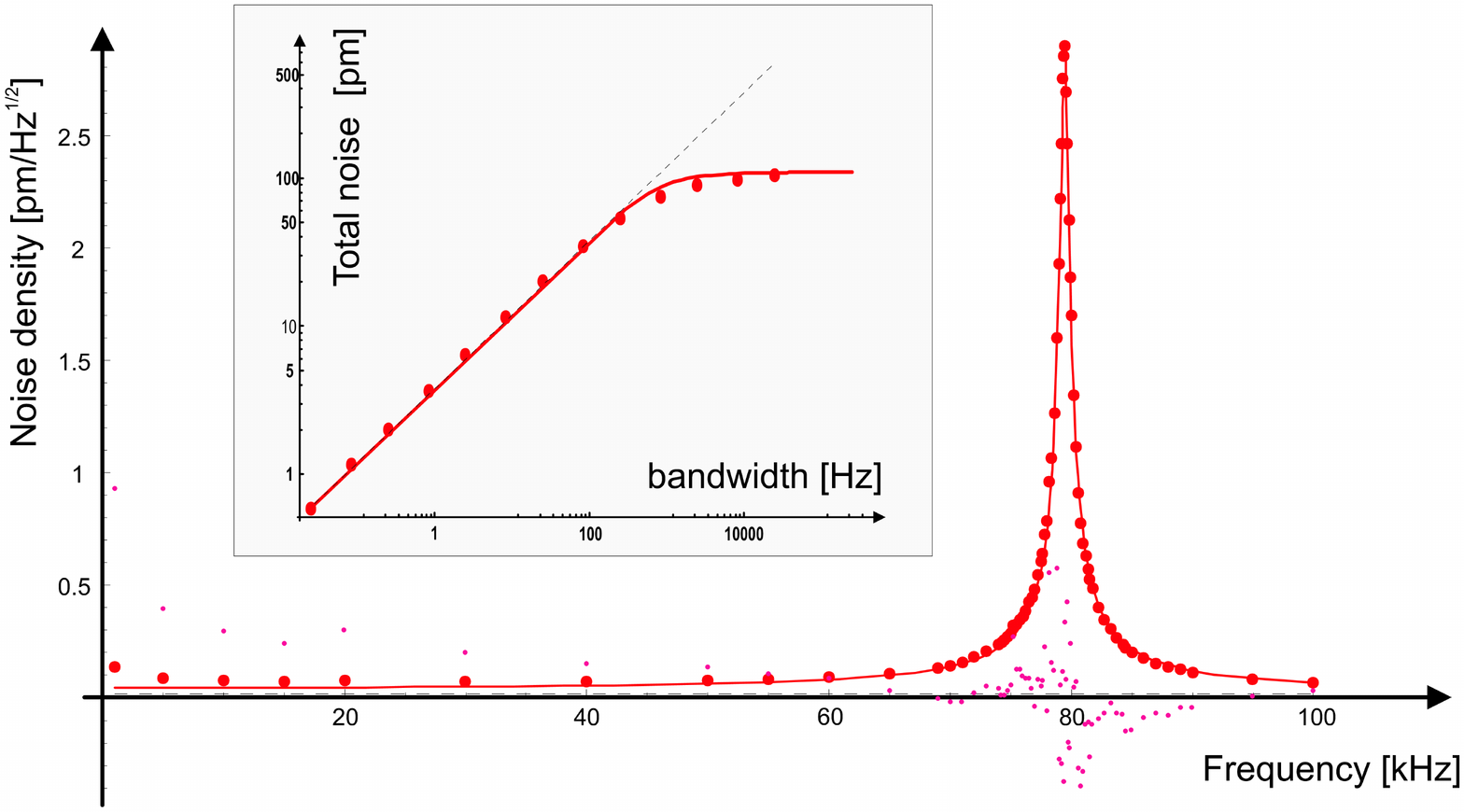}%
\caption{Main graph: Spectral noise density of a 0.4N/m cantilever
measured with a digital lock-in amplifier. For this noise
measurement, no external excitation was applied to the cantilever
and the motion of the cantilever was measured using the
beam-deflection technique. The larger (red) points correspond to
experimental noise data, the solid line to a fit assuming a constant
offset and a Lorenz function (see main text) and the smaller (pink)
points show the error between this fit and the measured data points.
Inset: Log-Log plot of the total noise as a function of bandwidth
for a noise measurement centered at the peak of the main noise
curve. For small bandwidth, the frequency noise shows the typical
$1/a_{os}$ behavior (slope -1 in the Log-Log plot). However, for
high bandwidth ($bw>\nu_{0}/Q$, with $Q$ quality factor) the total
noise saturates.}
\end{figure}

\begin{figure}[H]
\centering
\includegraphics[height=16cm,angle=-90]{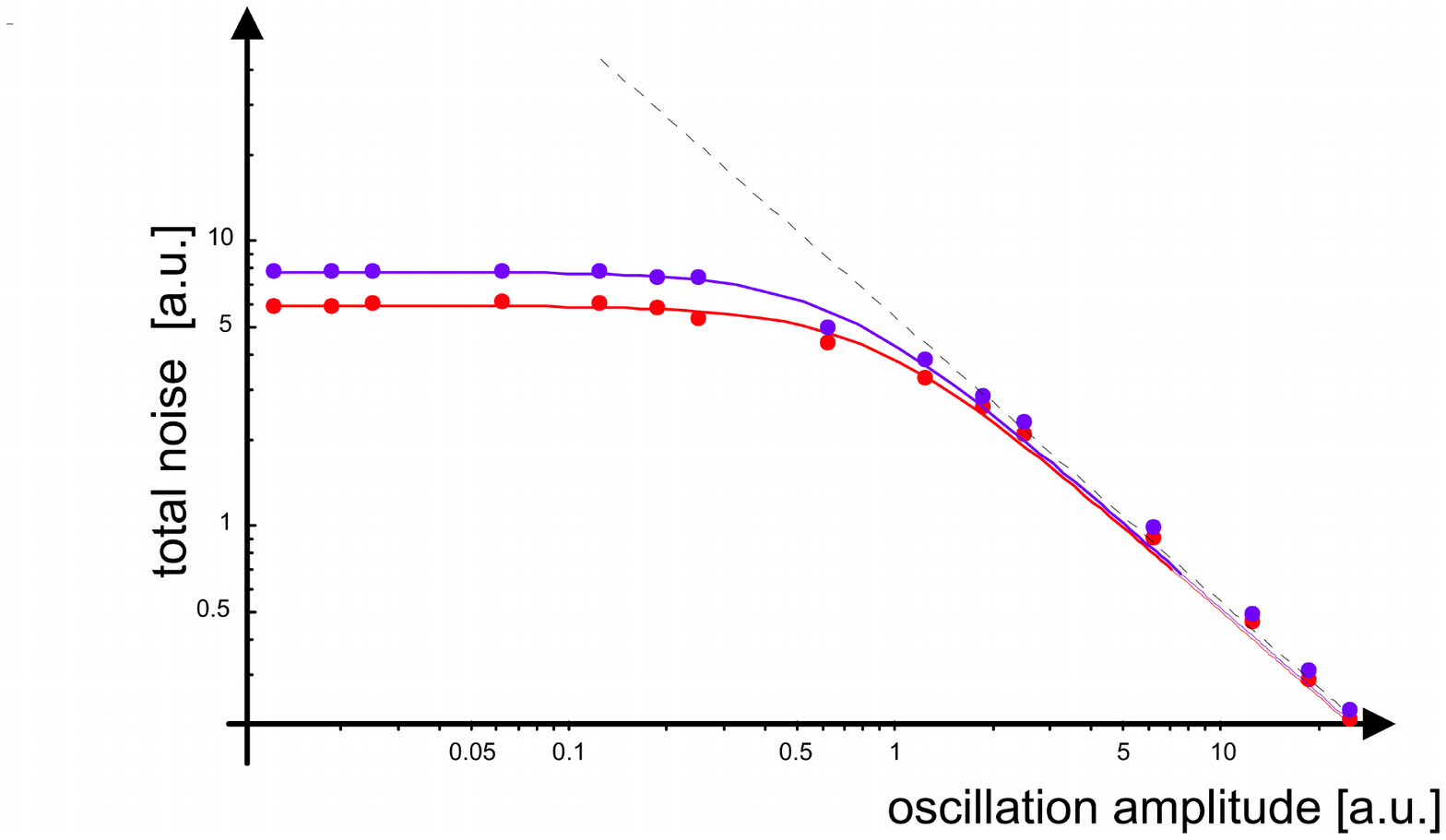}
\caption{Frequency noise of the DSFM\ detection electronics measured
as a function of oscillation amplitude for two different bandwidths
(50 Hz and 100 Hz). For this measurement, the same cantilever as
that used for the previous experiment was utilized (force constant
of 0.4N/m). The cantilever was excited by the DSFM electronics with
the phase-locked loop enabled, and the frequency output
$\delta\nu(t)$ was fed into a digital lock-in amplifier in order to
determine the total noise of the frequency measurement of the DSFM\
detection unit. For small oscillation amplitude of the cantilever
($a_{osci}<a_{th}$, see main text), the frequency noise is
independent of oscillation amplitude and for large amplitude the
noise decreases linearly (slope 1 in the $log-log$ plot).}
\end{figure}


\begin{thebibliography}{99}
\bibitem {CommentFreqAndOmega}We note that the different use of frequency
$\nu$ and angular frequency $\omega$ may lead to considerable confusion. In
the present work we will try to define all quantities in such a way that their
functional appearance is the same, independently of whether frequency or
angular frequency is used. Mostly frequency will be used, and the use of
angular frequency will be limited to terms like $\sin\left(  \omega t\right)
$ and $\cos\left(  \omega t\right)  $ where the factor $2\pi$ -as in
$\sin\left(  2\pi\nu t\right)  $-would be unnecessary bulky. In particular
spectral densities -as in \ref{densityNoise}- are referred to frequency and
not to angular frequency.
\bibitem {NotePhiIsZero}This may seem evident since we assume that the system
is at resonance. Mathematically, however, it is important that a finite
oscillation amplitude "defines" the mean angle: $\left\langle \varphi
(t)\right\rangle _{\tau}\simeq-\pi/2+\left\langle x(t)/y(t)\right\rangle
_{\tau}\simeq-\pi/2+\left\langle x(t)\right\rangle _{\tau}/\left\langle
y(t)\right\rangle _{\tau}=-\pi/2+0/\left(  Qa_{exc}\right)  =-\pi/2$, for
$a_{exc}\neq0$ , while for $a_{exc}=0$ this relation -and thus the mean angle-
is undefined.
\bibitem {CommentAmpContronNo}Note, however, that an optional oscillation
amplitude gain control, usually used in DSFM to measure dissipation, is not
included in figure 1. For the frequency noise measurements to be discussed
here the gain control of the DSFM electronics was disabled and is thus not
relevant in the present context.
\bibitem {CommentNoise}More precisely: only for constant noise density the
increase of total noise with will have a square root dependence. Note that if
the noise density has essentially two ranges, one where the thermal noise is
dominant and another one where a (constant) technical noise density is
relevant, in the log-log plot this will give rise to two regions, both with
slope 1/2 and thus parallel, but with a slight offset induced by the
transition from the frequency range dominated by thermal noise to the
frequency range dominated by technical noise.
\end{thebibliography}
\end{document}